\documentclass[%
 aip,%
 amsmath,amssymb,
reprint,%
]{revtex4-1}

\usepackage{graphicx}
\usepackage{dcolumn}
\usepackage{bm}

\usepackage{color}

\begin{document}

\preprint{AIP/123-QED}

\title{Exact axisymmetric Taylor states for shaped plasmas}

\author{Antoine J. Cerfon}
\email{cerfon@cims.nyu.edu}
\affiliation{ Courant Institute of Mathematical Sciences, New York University, New York, New York 10012}%
\author{Michael O'Neil}%
 \email{oneil@cims.nyu.edu}
\affiliation{ Courant Institute of Mathematical Sciences, New York University, New York, New York 10012}%

\date{\today}

\begin{abstract}
We present a general construction for exact analytic Taylor states in
axisymmetric toroidal geometries. In this construction, the Taylor equilibria are fully determined
by specifying the aspect ratio, elongation, and triangularity of the
desired plasma geometry. For equilibria with a magnetic X-point, the
location of the X-point must also be specified. The flexibility and
simplicity of these solutions make them useful for verifying the accuracy
of numerical solvers and for theoretical studies of Taylor states in
laboratory experiments.
\end{abstract}

\maketitle

Plasmas in both astrophysical and laboratory settings have a strong
tendency to relax to minimum energy states known as Taylor states or
Woltjer-Taylor states
\cite{Lust,Chandrasekhar,Woltjer,Taylor1,Taylor2,Shaffer,Geddes,Tang,Jarboe,Battaglia,Qin,Gray} in which the magnetic fields are force-free fields given by the equation
\begin{equation}\label{eq:force-free}
\nabla\times\mathbf{B}=\lambda \mathbf{B},
\end{equation}
where $\lambda$ is a global constant. A well-known analytic
solution to equation \eqref{eq:force-free} is often used for theoretical studies and to interpret experiments \cite{Taylor2, Shaffer, Geddes, Brown}. One of its main advantage is its simplicity, but it lacks the degrees of freedom necessary to describe the large variety of configurations observed in laboratory experiments. We present a new family of exact
solutions to equation \eqref{eq:force-free} and a general construction for the solutions that address this need. The new solutions, while still simple, have the flexibility to describe configurations within a wide range of aspect ratios, elongations, and triangularities. The plasma boundary can have a magnetic separatrix, if desired, and the location of the separatrix can be specified. The equilibria we describe in this article can thus be useful for a variety of applications, including the study of non-solenoidal current start-up in low aspect ratio toroidal
devices \cite{Battaglia} and plasma dynamics in spheromaks. They can also be used to verify the accuracy of
numerical schemes developed to solve equation \eqref{eq:force-free} in
fusion-relevant geometries \cite{Kress,epgron}. Efficient
solvers for force-free magnetic fields have recently become particularly attractive as a building
block in a promising formulation for three-dimensional equilibria in
fusion devices \cite{Dennis,Hudson}. The exact solutions we
present in this article can in that sense be thought of as the equivalent of
Solov'ev solutions used to benchmark Grad-Shafranov solvers which are
designed to compute more general equilibria \cite{Lutjens}. The ability
to construct exact equilibria with magnetic X-points is very desirable,
since X-points are usually a source of difficulty in both theoretical
studies and in numerical solvers.

Our construction of analytic solutions works as follows. We first turn
equation \eqref{eq:force-free} into its associated Grad-Shafranov
equation for the poloidal flux function $\psi$. We then express the
solution $\psi$ as a finite sum of functions satisfying the
Grad-Shafranov equation. Finally, in order to have the $\psi$ contours
conform with shaped plasmas relevant to laboratory experiments, we
determine the free constants appearing in the finite sum of functions
such that the edge of the plasma, given by the $\psi=0$ contour, is in
good agreement with a desired model surface, as was recently done for
Solov'ev profiles \cite{Cerfon}. The organization of the article follows the steps of the construction.

The current density in an axisymmetric toroidal geometry can be written
as \cite{Freidberg}
\begin{equation}\label{eq:current}
\begin{aligned}
\mu_{0}\mathbf{J}&=\mu_{0}(J_{T}\mathbf{e}_{\phi}+\mathbf{J}_{P}) \\
&=-\frac{1}{R}\Delta^{*}\psi\mathbf{e}_{\phi}+\frac{1}{R}\nabla
g\times\mathbf{e}_{\phi}
\end{aligned}
\end{equation}
where $\Delta^{*}$ is the operator
\begin{displaymath}
\Delta^{*}\equiv R\frac{\partial}{\partial
  R}\left(\frac{1}{R}\frac{\partial\psi}{\partial
  R}\right)+\frac{\partial^{2}\psi}{\partial Z^{2}},
\end{displaymath}
$(R,\phi,Z)$ is the natural cylindrical coordinate system associated
with the toroidal geometry, $\mathbf{e}_{\phi}$ is the unit vector in
the toroidal direction $\phi$, $2\pi\psi(R,Z)$ is the poloidal
magnetic flux, $2\pi g(\psi)=-I_{p}(\psi)$ is the net poloidal current
flowing in the plasma and the toroidal field coils, the letter T
stands for toroidal, and the letter P stands for poloidal. The magnetic field is then given by
\begin{equation}
\begin{aligned}
\mathbf{B}&=B_{T}\mathbf{e}_{\phi}+\mathbf{B}_{P}\\
&=\frac{g(\psi)}{R}\mathbf{e}_{\phi}+\frac{1}{R}\nabla\psi\times\mathbf{e}_{\phi}.
\end{aligned}
\end{equation}

A Taylor state satisfies the condition
$\mu_{0}\mathbf{J}=\lambda\mathbf{B}$, which implies in the toroidal
and poloidal directions:
\begin{equation}\label{eq:system}
\begin{aligned}
 -\frac{1}{R}\Delta^{*}\psi&=\lambda\frac{g(\psi)}{R},\\
 \frac{1}{R}\frac{dg}{d\psi}&=\frac{\lambda}{R}.
\end{aligned}
\end{equation}
The second equation of the system can be easily integrated, and we
find that
\begin{equation}\label{eq:poloidal_cur}
g(\psi)=\lambda\psi,
\end{equation}
where the free constant of integration is set to zero to correspond to
a situation with no vacuum toroidal field. Using expression
\eqref{eq:poloidal_cur} for $g(\psi)$ in the first equation of system
\eqref{eq:system}, we obtain the desired Grad-Shafranov equation
corresponding to Taylor states:
\begin{equation}\label{eq:gradshaf}
\Delta^{*}\psi=-\lambda^{2}\psi.
\end{equation}

We now construct an analytic solution $\psi$ to the following problem:
\begin{equation}\label{eq:GS_with_bd}
\begin{aligned}
 \Delta^{*}\psi&=-\lambda^{2}\psi & \qquad &\text{in } \Omega,\\
 \psi&=0 & &\text{on } \partial\Omega,
\end{aligned}
\end{equation}
in which the domain $\Omega$ is relevant to axisymmetric toroidal plasma
experiments. We do this by constructing a solution that has enough
degrees of freedom to satisfy the condition $\psi=0$ at a few points
on a model surface \cite{Cerfon}, and by defining, after the fact,
$\partial \Omega$ by the implicit equation
$\psi(R,Z)=0$. Even though \eqref{eq:GS_with_bd} has
$\psi\equiv0$ as a trivial solution, our procedure avoids it and solves for the desired Taylor state.

Let us first focus on the equation 
\begin{equation}\label{eq:homog}
\Delta^{*}\psi=-\lambda^{2}\psi
\end{equation}
without concern for the boundary
conditions. Equation~\eqref{eq:homog} can be solved via separation of variables \cite{Chandrasekhar}. Writing
$\psi(R,Z)=F(R)H(Z)$, we have
\begin{multline}\label{eq:separated}
H(Z)\frac{d^{2}F}{dR^{2}}-\frac{H(Z)}{R}\frac{dF}{dR}+F(R)\frac{d^{2}H}{dZ^{2}}\\ 
=-\lambda^{2}F(R)H(Z).
\end{multline}
Setting
\begin{equation}\label{eq:H_eq}
\frac{d^{2}H}{dZ^{2}}=-k^{2}H(Z),
\end{equation}
equation \eqref{eq:separated} becomes
\begin{equation}\label{eq:F_radial}
\frac{d^{2}F}{dR^{2}}-\frac{1}{R}\frac{dF}{dR}+(\lambda^{2}-k^{2})F(R)=0.
\end{equation}
For $\lambda^{2}\geq k^{2}$, the general solution to this equation is
\cite{McCarthy}
\begin{multline}
F(R)=R\left[c \, J_{1}\left(\sqrt{\lambda^{2}-k^{2}}R\right) \right.\\
+\left. d \, Y_{1}\left(\sqrt{\lambda^{2}-k^{2}}R\right)\right],
\end{multline}
where $J$ is the Bessel function of the first kind, $Y$ the Bessel
function of the second kind, and $c$ and $d$ are constants.

The solution of equation \eqref{eq:H_eq} is
\begin{equation}
H(Z)=e \, \cos(kZ)+f \, \sin(kZ),
\end{equation}
where $e$ and $f$ are constants. Note finally that there exists
another type of solution to equation \eqref{eq:homog}:
\begin{equation}
\psi(R,Z) = \cos\left(\lambda\sqrt{R^{2}+Z^{2}}\right).
\end{equation}

As we will see next, in the case of up-down asymmetric equilibria, we
will impose twelve boundary conditions on the general solution in
order to have optimal agreement between the desired boundary
and the implicit boundary $\partial\Omega$ given by $\psi(R,Z)=0$. We
therefore choose the following general solution with twelve degrees of
freedom:
\begin{align}\label{eq:general_sol}
&\psi(R,Z,c_{1},c_{2},c_{3},c_{4},c_{5},c_{6},c_{7},c_{8},c_{9},c_{10},c_{11},c_{12})=\psi_{0}\notag\\
&+c_{1}\psi_{1}+c_{2}\psi_{2}+c_{3}\psi_{3}+c_{4}\psi_{4}\notag\\
&+c_{5}\psi_{5}+c_{6}\psi_{6}+c_{7}\psi_{7}\notag\\
&+c_{8}\psi_{8}+c_{9}\psi_{9}+c_{10}\psi_{10}
\end{align}
with
\begin{align*}
&\psi_{0}=RJ_{1}(c_{12}R)\;,\;\;\psi_{1}=RY_{1}(c_{12}R)\\
&\psi_{2}=RJ_{1}\left(\sqrt{c_{12}^{2}-c_{11}^{2}}R\right)\mbox{cos}(c_{11}Z)\\
&\psi_{3}=RY_{1}\left(\sqrt{c_{12}^{2}-c_{11}^{2}}R\right)\mbox{cos}(c_{11}Z)\\
&\psi_{4}=\mbox{cos}\left(c_{12}\sqrt{R^{2}+Z^{2}}\right)\;,\;\;\psi_{5}=\mbox{cos}\left(c_{12}Z\right)\\
&\psi_{6}=RJ_{1}(c_{12}R)Z\;,\;\;\psi_{7}=RY_{1}(c_{12}R)Z\\
&\psi_{8}=RJ_{1}\left(\sqrt{c_{12}^{2}-c_{11}^{2}}R\right)\mbox{sin}(c_{11}Z)\\
&\psi_{9}=RY_{1}\left(\sqrt{c_{12}^{2}-c_{11}^{2}}R\right)\mbox{sin}(c_{11}Z)\\
&\psi_{10}=\mbox{sin}(c_{12}Z)
\end{align*}
and where $k=c_{11}$ and $\lambda=c_{12}$ are treated as unknowns. The
twelve unknowns $c_{1},\ldots,c_{12}$ are obtained by specifying boundary conditions. We now explain how to do so for plasma equilibria in laboratory experiments.

We specify the unknowns $c_{1},\ldots,c_{12}$ so
as to best approximate the plasma boundary of interest. As an illustration, consider the following parametric curve, which describes a wide class of experimentally relevant axisymmetric plasma boundaries
\cite{Cerfon,Miller},
\begin{equation}\label{eq:boundary}
\begin{aligned}
 R(t)&=1+\epsilon  \cos\left(t+\alpha  \sin t \right)\\
 Z(t)&=\epsilon \kappa  \sin t,
\end{aligned}
\end{equation}
for $0\leq t < 2\pi$. $\epsilon$ is the inverse aspect ratio, $\kappa$
is the elongation, and $\sin \alpha=\delta$ is the
triangularity. In terms of these parameters, the outer equatorial
point has coordinates $(1+\epsilon,0)$, the inner equatorial point has
coordinates $(1-\epsilon,0)$, and the bottom point has coordinates
$(1-\delta\epsilon,-\kappa\epsilon)$. We will also need the curvatures
at these three points, given by:
\begin{equation}
N_{1}=-\frac{(1+\alpha)^2}{\epsilon\kappa^{2}}\;,\; N_{2}=\frac{(1-\alpha)^2}{\epsilon\kappa^{2}}\;,\;N_{3}=\frac{\kappa}{\epsilon\mbox{cos}^{2}\alpha}.
\end{equation} 
The geometric constraints imposed to determine the coefficients
$c_{1},\ldots,c_{12}$ for up-down {\em asymmetric} equilibria with a
magnetic separatrix are as follows \cite{Cerfon}:
\begin{equation}\label{eq:constraints}
\begin{cases}
\psi(1+\epsilon,0,C)=0\\
\psi(1-\epsilon,0,C)=0\\
\psi(1-\delta\epsilon,-\kappa\epsilon,C)=0\\
\psi_{R}(1-\delta\epsilon,-\kappa\epsilon,C)=0\\
\psi_{ZZ}(1+\epsilon,0,C)+N_{1}\psi_{Z}(1+\epsilon,0,C)=0\\
\psi_{ZZ}(1-\epsilon,0,C)+N_{2}\psi_{Z}(1-\epsilon,0,C)=0\\
\psi_{RR}(1-\delta\epsilon,-\kappa\epsilon,C)+N_{3}\psi_{Z}(1-\delta\epsilon,-\kappa\epsilon,C)=0\\
\psi(R_{sep},Z_{sep},C)=0\\
\psi_{R}(R_{sep},Z_{sep},C)=0\\
\psi_{Z}(R_{sep},Z_{sep},C)=0\\
\psi_{Z}(1+\epsilon,0,C)=0\\
\psi_{Z}(1-\epsilon,0,C)=0
\end{cases}
\end{equation}
where $C=(c_{1},\dots,c_{12})$, $(R_{sep}, Z_{sep})$ are the
coordinates of the magnetic X-point, and the subscripts refer to
partial derivatives with respect to the specified variable. The first
three conditions specify the location of the outer equatorial point,
inner equatorial point, and bottom point of the plasma boundary. The
fourth condition guarantees that the normal component of the poloidal
field is zero at the bottom point. The fifth, sixth, and seventh
conditions determine the local curvature of the plasma boundary at the
outer equatorial point, inner equatorial point, and bottom point,
respectively. The eigth, ninth and tenth conditions impose the
presence of a magnetic X-point at $(R_{sep},Z_{sep})$ on the plasma
boundary. The last two conditions give the slope of the plasma
boundary at the outer equatorial point and inner equatorial
point. This is necessary for equilibria that are not up-down symmetric.

Equation \eqref{eq:constraints} is a non-linear system of 12 equations
for 12 unknowns. Given good initial conditions, it can be solved
without difficulty using standard non-linear root finding packages, such as \texttt{fsolve} in MATLAB
\cite{Matlab}. Solutions were found to an absolute precision of at
  least $10^{-16}$ in all of the examples shown in this article. An efficient way
to get a good initial guess is to first treat $c_{11}$ and $c_{12}$ as
constants and solve what then becomes a linear system. In order to
solve the system \eqref{eq:constraints}, it is best to use exact
formulae for $\psi_{R}$, $\psi_{RR}$, $\psi_{Z}$, and $\psi_{ZZ}$. The
calculation of these partial derivatives is straight-forward. For the
$R$ derivatives of the terms involving Bessel functions, the following formulae, valid for
an arbitrary real constant $\mu$ and obtained from Bessel identities, are
useful:
\begin{align*}
\frac{d}{dR}\left(RJ_{1}(\mu R)\right)&=J_{1}(\mu
  R)+\frac{\mu}{2}R\left[J_{0}(\mu R)-J_{2}(\mu
    R)\right]\\ 
\frac{d}{dR}\left(RY_{1}(\mu R)\right)&=Y_{1}(\mu
  R)+\frac{\mu}{2}R\left[Y_{0}(\mu R)-Y_{2}(\mu
    R)\right]\\ 
\frac{d^{2}}{dR^{2}}\left(RJ_{1}(\mu
  R)\right)&=\left(\frac{1}{R}-\mu^{2}R\right)J_{1}(\mu
  R)\\ 
  &\qquad\qquad+\frac{\mu}{2R}\left[J_{0}(\mu
    R)-J_{2}(\mu R)\right]\\ 
\frac{d^{2}}{dR^{2}}\left(RY_{1}(\mu
  R)\right)&=\left(\frac{1}{R}-\mu^{2}R\right)Y_{1}(\mu
  R)\\ 
  &\qquad\qquad+\frac{\mu}{2R}\left[Y_{0}(\mu
    R)-Y_{2}(\mu R)\right]
\end{align*}

Assuming we have calculated $\psi$ according to this
procedure, consider the magnetic field $\mathbf{B}$ given by
\begin{equation}\label{eq:magnetic_field}
\mathbf{B}=B_{\phi}\mathbf{e}_{\phi}+\mathbf{B}_{p}=\frac{c_{12}\psi}{R}+\frac{1}{R}\nabla\psi\times\mathbf{e}_{\phi},
\end{equation}
and the toroidal flux
\begin{displaymath}
\Psi=c_{12}\iint_{\Omega} \frac{\psi}{R} \, dR \, dZ,
\end{displaymath}
where $\Omega$ is the region inside of the boundary $\partial\Omega$ given by the implicit equation
$\psi(R,Z)=0$. The magnetic field $\mathbf{B}$ as defined in
equation \eqref{eq:magnetic_field} is in the axisymmetric Taylor
state described by:
\begin{equation}
\begin{aligned}
\nabla\times\mathbf{B}&=c_{12}\mathbf{B}\\
\iint_{\Omega}\mathbf{B}\cdot d\mathbf{S}&=\Phi.
\end{aligned}
\end{equation}

The method we present in this article leads to $c_{12}\geq0$ and
$\psi\geq0$, corresponding to right-handed Taylor states
\cite{Brown}. Left-handed Taylor states can be constructed from these
solutions without difficulty. Indeed, if we define $\varphi=-\psi$ and
$\gamma=-c_{12}$, then it is easy to see that the magnetic field
$\mathbf{B}_{L}$ defined by
\begin{equation}\label{eq:left_handed}
\mathbf{B}_{L}=\frac{\gamma\varphi}{R}+\frac{1}{R}\nabla\varphi\times\mathbf{e}_{\phi}
\end{equation}
is in the left-handed Taylor state given by:
\begin{equation}
\begin{aligned}
\nabla\times\mathbf{B}_{L}&=\gamma\mathbf{B}_{L}\\
\iint_{\Omega}\mathbf{B}_{L}\cdot d\mathbf{S}&=\Phi.
\end{aligned}
\end{equation}

\begin{figure}
\center
\includegraphics[width=0.9\linewidth]{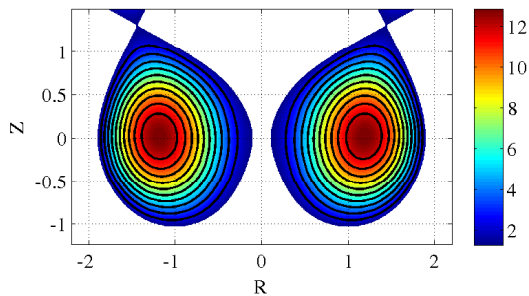}
\caption{Contours of the poloidal flux $\psi$ for $\epsilon=0.9$, $\kappa=1.15$, $\delta=0$, and $(R_{sep},Z_{sep})=(1+\epsilon/2,5\epsilon\kappa/4)$, in arbitrary units.}
\label{fig:asym_contour_plot}
\end{figure}

\begin{figure}
\center
\includegraphics[width=0.9\linewidth]{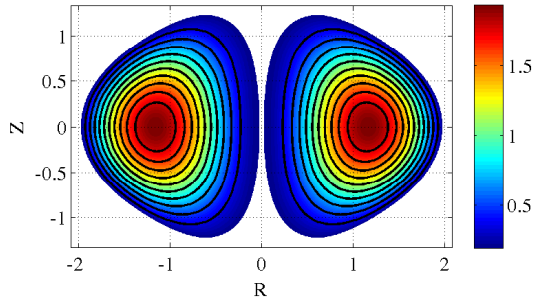}
\caption{Contours of the poloidal flux $\psi$ for $\epsilon=0.98$,
  $\kappa=1.25$ and $\delta=0.4$, in arbitrary units.}
\label{fig:sym_contour_plot}
\end{figure}

We have found empirically that our general construction of Taylor states is very robust, leading to physically relevant equilibria over a wide range of aspect ratios, elongations, triangularities, and locations of the magnetic X-point. We show two examples illustrating this point. Figure \ref{fig:asym_contour_plot} is a contour plot of the flux function $\psi$ for an up-down asymmetric Taylor state with a magnetic X-point on the plasma boundary, and geometric parameters $\epsilon=0.9$, $\kappa=1.15$, $\delta=0$, and $(R_{sep},Z_{sep})=(1+\epsilon/2,5\epsilon\kappa/4)$. The second example is an up-down symmetric equilibrium. Such equilibria can be constructed using the same
procedure as the one for asymmetric equilibria: one sets $c_{6}=\ldots=c_{10}=0$, strips the system
\eqref{eq:constraints} of the last 5 equations, and solves the
remaining non-linear system of 7 equations for the seven unknowns to
compute $c_{1}$, $c_{2}$, $c_{3}$, $c_{4}$, $c_{5}$, $c_{11}$,
$c_{12}$. Figure \ref{fig:sym_contour_plot} is a contour plot of $\psi$ for an up-down symmetric Taylor state with geometric parameters $\epsilon=0.98$, $\kappa=1.25$, and $\delta=0.4$

\begin{figure}
\center
\includegraphics[width=0.75\linewidth]{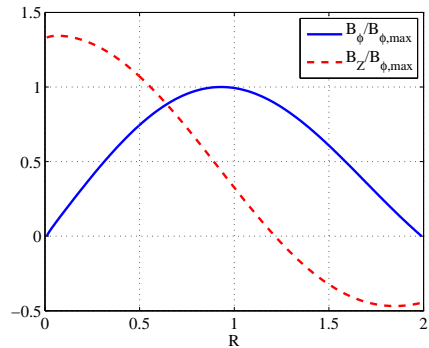}
\caption{Toroidal (blue continuous line) and poloidal (red dashed
  line) magnetic field at $Z=0$ for $\epsilon=0.99$, $\kappa=1.22$ and
  $\delta=0$ for a right-handed orientation. The fields have been
  normalized to the maximum of the toroidal field.}
\label{fig:fields}
\end{figure}

The equilibria we present in this article are a good approximation of experimental observations and numerical simulations of force-free equilibria in laboratory experiments. As an illustration, we plot in Figure \ref{fig:fields} the toroidal and poloidal magnetic fields at the midplane $Z=0$, normalized to the maximum of the toroidal field, for parameters relevant to the Swarthmore Spheromak Experiment (SSX) \cite{Geddes}: $\epsilon=0.99$, $\kappa=1.22$, $\delta=0$. Comparing Figure \ref{fig:fields} with Figure 7 in Reference 7, one can see that the magnetic field profiles agree well, both in terms of shape and relative magnitude, with those observed during the early decay phase in SSX, which corresponds to the constant $\lambda$ phase. For $\epsilon=0.99$, $\kappa=1.22$, $\delta=0$ we obtain $\lambda\approx 19.6\;m^{-1}$, to be compared with the numerically computed value $\lambda\approx18.4\;m^{-1}$. This discrepancy can be explained by the fact that the parametric equations \eqref{eq:boundary} describe a surface that is smoother than the rectangular flux conserver in SSX. If better quantitative agreement is desired, \eqref{eq:boundary} and \eqref{eq:constraints} can be readily modified to better conform to the specific geometry of interest. 

In summary, we have presented the first explicit construction of toroidally
axisymmetric Taylor states with boundary conditions relevant to shaped
plasmas in laboratory experiments. In this construction, the Taylor
states are expressed in terms of the poloidal magnetic flux function
$\psi$ which is described by the sum of at most 12 terms, all of which are simple functions of $R$,
$Z$ with explicit derivatives of any order. Despite their simplicity,
the Taylor equilibria we present are very versatile. They can be used
to describe plasma boundaries with or without a magnetic $X$-point,
and with a wide range of aspect ratios, elongations, and
triangularities. They are therefore useful for a variety of
applications, such as theoretical studies of Taylor states in very low aspect ratio experiments and benchmarking the accuracy of numerical solvers for force-free magnetic fields.

This research was supported in part by the U.S. Department of Energy, Office of Science, Fusion Energy
Sciences under award number DE-FG02-86ER53223 (A. Cerfon) and in part by the Air Force Office of Scientiﬁc Research under NSSEFF Program Award FA9550-10-1-0180 (M. O'Neil).

\end{document}